\documentclass{elsart}
\usepackage{epsfig}
\begin{document}
\begin{frontmatter}
\title{Hints for an extension of the early exercise premium formula for American options}
\author[lu]{Hans-Peter Bermin},
\author[po]{Arturo Kohatsu-Higa}, \and
\author[ub]{Josep Perell\'o}
\ead{josep.perello@ub.edu}
\address[lu]{Product Development Global Financial Markets, WestLB Securities Pacific Ltd., Tokyo Branch Roppongi Hills Mori Tower 37th Floor, 10-1 Roppongi 6-chome Minato-ku, Tokyo 106-6137, Japan}
\address[po]{Department of Economics, Universitat Pompeu Fabra, Ramon Trias Fargas 25-27, 08005-Barcelona, Spain}
\address[ub]{Departament de F\'{\i}sica Fonamental, Universitat de Barcelona, Diagonal 647, 08028-Barcelona, Spain}
\maketitle
\begin{abstract}
Characterization of the American put option price is still an open issue. From the beginning of the nineties there exists a non-closed formula for this price but nontrivial numerical computations are required to solve it. Strong efforts have been done to propose methods more and more computationally efficient but most of them have few mathematical ground as to ascertain why these methods work well and how important is to consider a good approximation to the boundary or to the smooth pasting condition. We perform an extension of the American put price aiming to catch weaknesses of the numerical methods given in the literature.
\end{abstract}
\end{frontmatter}
In this paper we study approximations for the price of the American put option. In recent years there has been an explosion of literature regarding this problem. The mathematical literature has focussed on the study of properties of the so-called optimal boundaries and the convergence properties of the binomial method of pricing (see Kim~\cite{kim}, Jacka~\cite{jacka} and Carr~\cite{carr}). In the more numerical literature, on the other hand, the main objective is to achieve a fast calculation of put prices, somewhat disregarding the approximation of the optimal boundary and the so-called smooth pasting condition. Huang {\it et al.} wrote: ``It seems that in order to have a good approximation of the price is not necessary to have a good approximation of the optimal boundary"~\cite{huang}. Subsequent works have improved their methodologies --see for instance Refs.~\cite{ju,aitsahlia}--. A common feature though among the computational articles is that they never investigate why a proposed method work well or not, and how important is to accurately approximate the optimal boundary. Aiming to answer these questions, our objective is to bridge the gap between these computational investigations and the more theoretical literature with the hope to bring some new insights. We must admit that this is a difficult task and the present study is limited to provide few insights. At this step, we solely sketch some results of a more extensive work which is still in progress~\cite{bermin}.

Let us provide the general framework. We consider the classical Black-Scholes model consisting of two assets: one risk free bond $B$ whose dynamics is described by 
\begin{equation}
dB\left( t\right) = rB\left( t\right) dt \qquad\mbox{with } B\left(0\right) =1,
\label{bond}
\end{equation}
and one stock of price $S$ following the stochastic differential equation\footnote{For sake of simplicity, in this paper we assume no dividend yield although similar conclusions can be obtained with non-zero dividend yield~\cite{bermin}.}
\begin{equation}
dS(t)=rS(t)dt+\sigma S(t)dW(t) \qquad \mbox{with } S(0) =S_0,
\label{stock}
\end{equation} 
and where $dW(t)=\xi(t)dt$ is the Wiener process with zero mean and unit variance. 

The American put option gives to its owner the right (and not the obligation) to sell an asset at a given strike price $K$ at any time up until maturity $T$. This last condition makes the pricing problem very complicated, much more than for European options where exercising the option is only possible at maturity date $T$. Under the Black-Scholes model determined by the pair of Eqs.~(\ref{bond}) and~(\ref{stock}), the American put price at time $t$ with strike price $K$ is defined by
\begin{equation}
p^{A}(t,S) \equiv\sup_{\tau \in {\mathcal S}_{t,T}}e^{-r(\tau-t)}\left\langle[K-S(\tau)]^{+}|S(t)=S\right\rangle,
\label{americandef}
\end{equation}
where $\mathcal S_{t,T}$ is the set of all stopping times with values in $[t,T]$. This definition involves an expectation over the put payoff, which solely takes the positive contribution of the difference between strike $K$ and stock $S$. We will fix the fair price as the highest average of the collection of expectations with stopping times $\tau \in [t,T]$. Hence, the problem reduces to an optimal-stopping-time problem and main purpose is to know the best moment to exercise the American option.

The pricing problem was solved more than a decade ago departing from discrete time and recursively backward with non arbitrage arguments~\cite{kim}. The so-called binomial pricing method gives a non-closed solution for the American put option in terms of the optimal exercise boundary $c(t)$ and reads (for alternative derivations see Refs.~\cite{jacka,carr})
\begin{eqnarray}
p^{A}(t,S)=p^{E}(t,S)&+&rK\int_{t}^{T}e^{-r(u-t)}N\left[-d_2(S,c(u),u-t)\right] du
\label{american}
\end{eqnarray}
where
\begin{eqnarray}
p^E(t,S)=Ke^{-r(T-t)}N[-d_2(S,K,T-t)]-SN[-d_1(S,K,T-t)]
\label{europut}
\end{eqnarray}
is the European put option price and 
\begin{equation}
N(a)\equiv\frac{1}{\sqrt{2\pi}}\int_{-\infty}^{a}e^{-x^2/2}dx
\label{N}
\end{equation}
is the probability integral function whose arguments are
\begin{eqnarray}
&&d_1(x,y,t)\equiv\frac{\ln(x/y)+(r+\sigma^2/2)t}{\sigma\sqrt{t}},
\qquad
d_2=d_1-\sigma\sqrt{t}.
\label{d}
\end{eqnarray}
The formula~(\ref{american}) says that the price of the American put is the same as the European put but with two extra contributions quantifying the premium associated to having the right of exercising early the option. In the put case, the optimal boundary characterizes the values below which we should exercise the American option ({\it i.e.}, if $S(t)<c(t)$) and it is solution to the equation:
\begin{eqnarray}
K-c(t)=p^E[t,c(t)]&+&rK\int_t^T e^{-r(u-t)}N[-d_2(c(t),c(u),u-t)]du.
\label{optimal}
\end{eqnarray}
This integral equation uniquely characterizes the optimal boundary among all non-decreasing continuous functions with values in $(0,K]$~\cite{jacka} and jointly with Eq.~(\ref{american}) constitute the non-closed formula for the American put price. Still, there is little hope of finding an analytical and explicit solution to Eq.~(\ref{optimal}) and subsequently to Eq.~(\ref{american}). For practical purposes though it gives us the possibility to numerically generate the optimal boundary, from which an approximate put price and replicating strategy can be derived. Each method will have their own clever tricks in order to make more efficient the computation.
\begin{figure}
\begin{center}
\epsfig{file=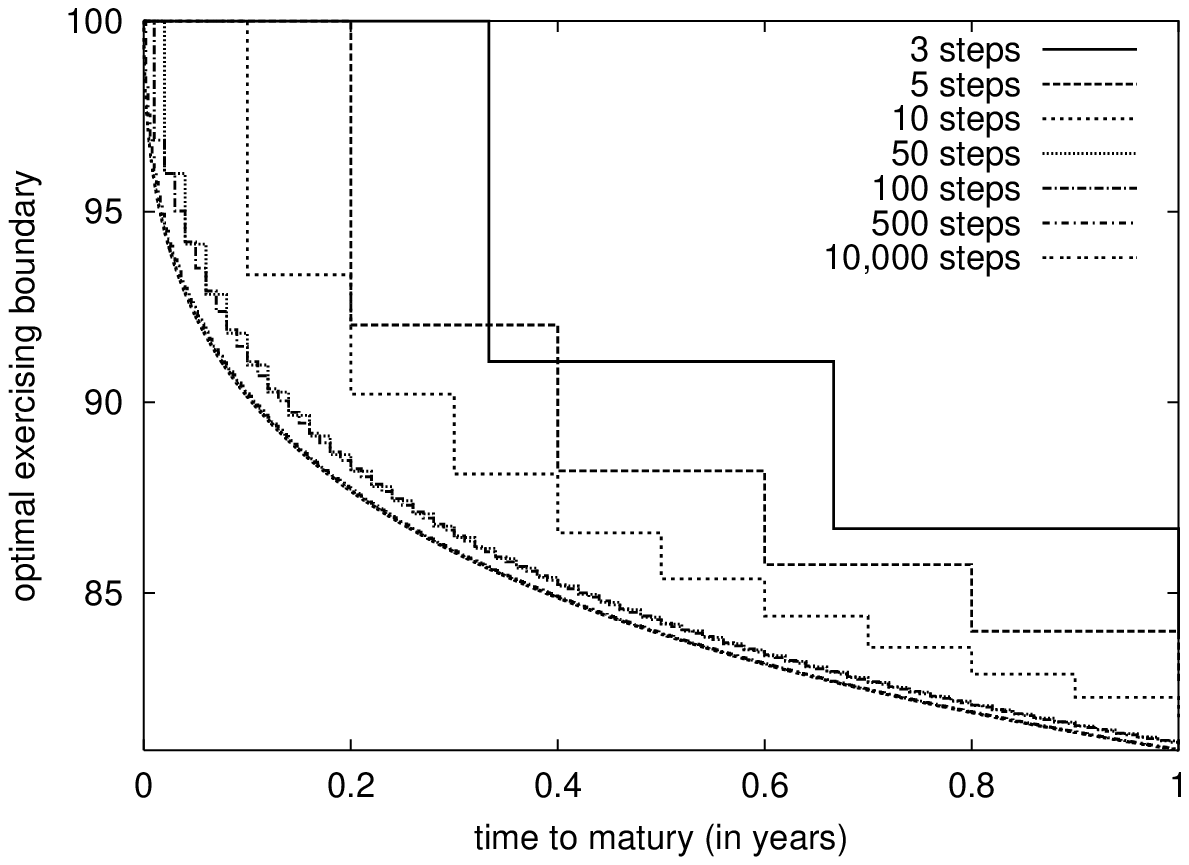,scale=0.54}
\epsfig{file=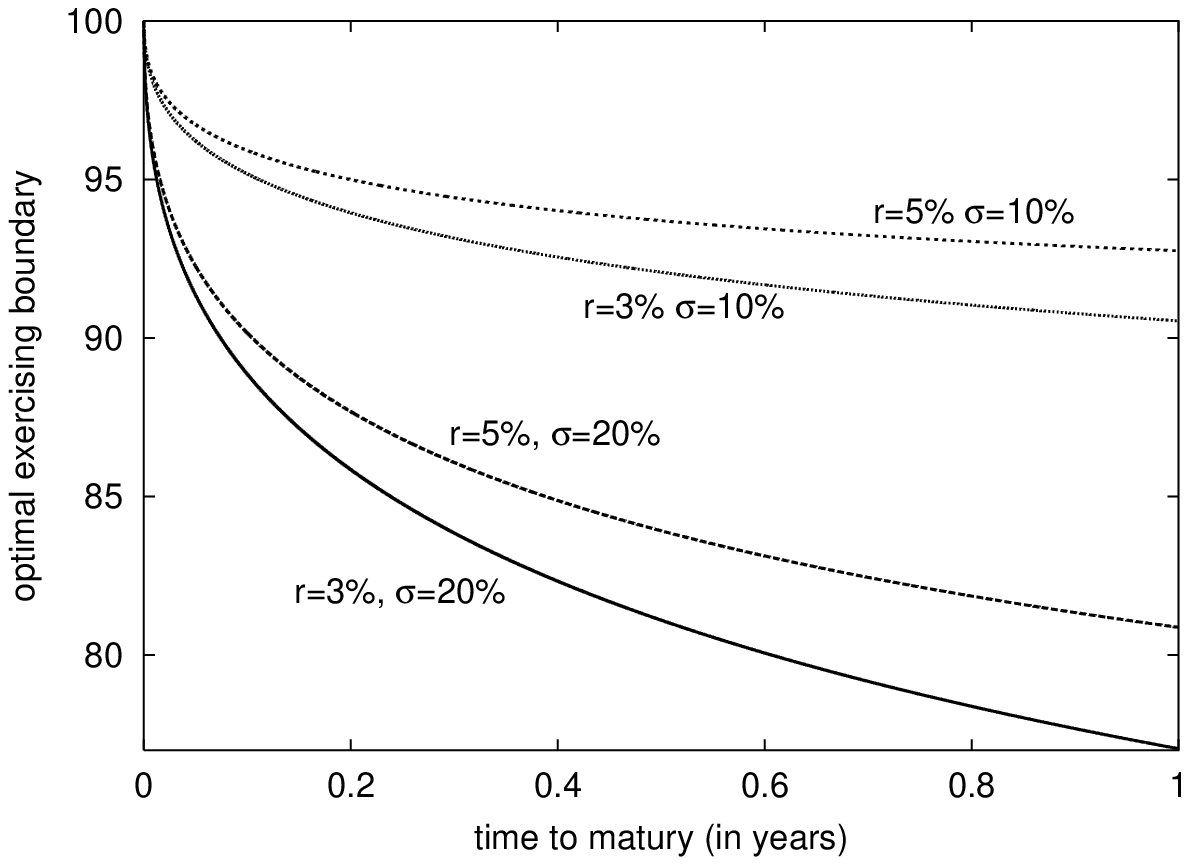,scale=0.54}
\end{center}
\caption{Computation of the optimal exercise boundary in terms of $T-t$. Eq.~(\ref{optimal}) is taken in its discrete form assuming that the optimal curve follows a stepped function and with $K=100$. Left hand side plot shows the quick convergence of the optimal curve when $r=5\%$ and $\sigma=20\%$. No big differences exist between the case of $N=500$ steps and the $N=10,000$ case. Right hand side shows the results for a set of $r$'s, and $\sigma$'s with $N=500$ steps. Both graphs take time units in years.}
\label{fig1}
\end{figure}

An alternative method for numerical computation arises from the Black-Scholes partial differential equation (PDE) approach, an approach also known as the Stefan problem. Taking finite differences, this method asserts that when underlying is above the optimal exercise boundary ($x>c(t)$) the PDE reads~\cite{merton}
\begin{equation}
\frac{\partial p^{A}(t,x)}{\partial t}+\frac{\sigma ^{2}x^{2}}{2}\frac{\partial ^{2}p^{A}(t,x)}{\partial x^{2}}+rx\frac{\partial p^{A}(t,x)}{\partial x}-rp^{A}(t,x)=0,
\label{pde}
\end{equation}
while below the boundary ($0<x\leq c(t)$) we immediately exercise the option with benefit 
\begin{equation}
p^{A}(t,x)=K-x.
\label{below}
\end{equation}
For continuity and theoretical arguments~\cite{merton}, both price and first order derivative over the price ($p^A$ and $p_x^A$) must be continuous on the boundary curve $c(t)$. In particular, last requirement gives the so-called smooth pasting condition. That is:
\begin{equation}
p_x^{A}[t,c(t)]\equiv\left.\frac{\partial p^{A}(t,x)}{\partial x}\right|_{x=c(t)}=-1.
\label{smooth}
\end{equation}
It is somewhat deceiving that the pricing method mentioned above~\cite{kim} does not clearly show how the smooth pasting condition~(\ref{smooth}) is incorporated into early exercise premium formula~(\ref{american}). This is a serious drawback of the numerical methods which are based on premium formula~(\ref{american}). 
\begin{figure}
\begin{center}
\epsfig{file=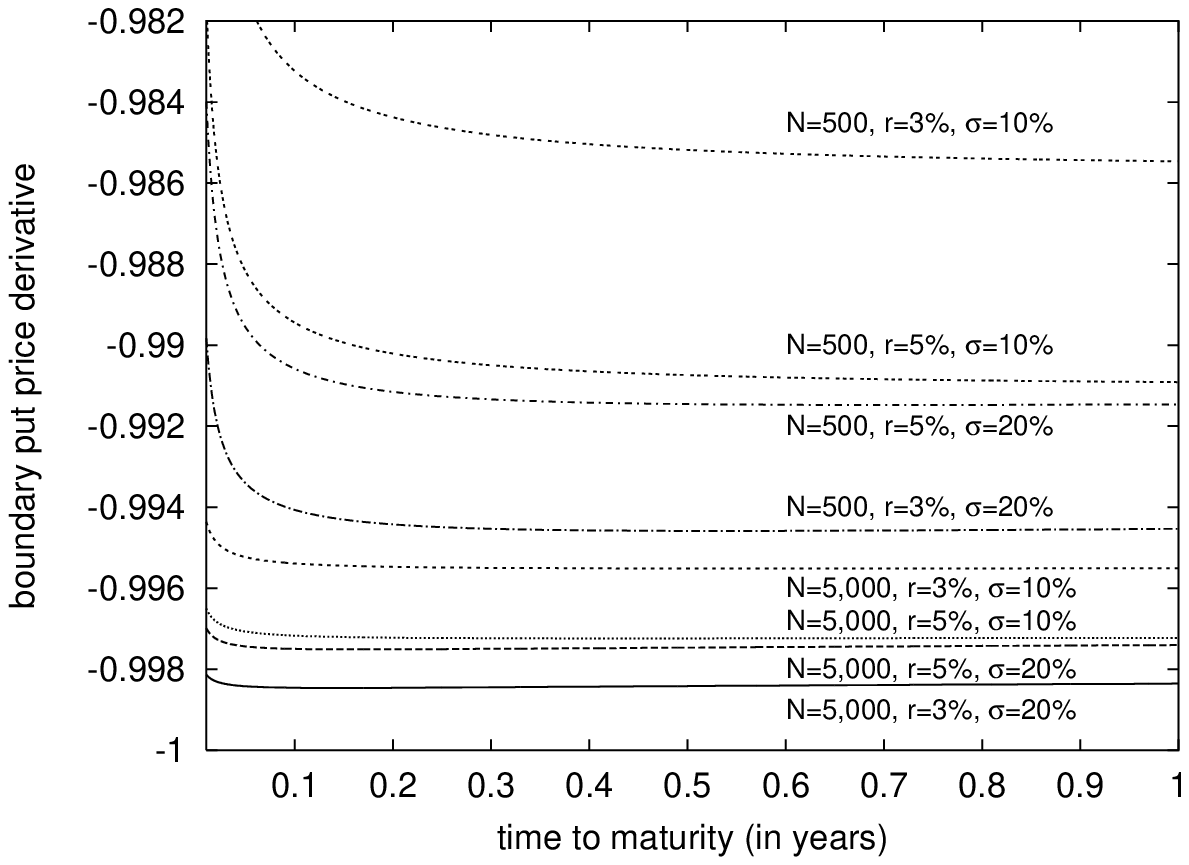,scale=0.54}
\epsfig{file=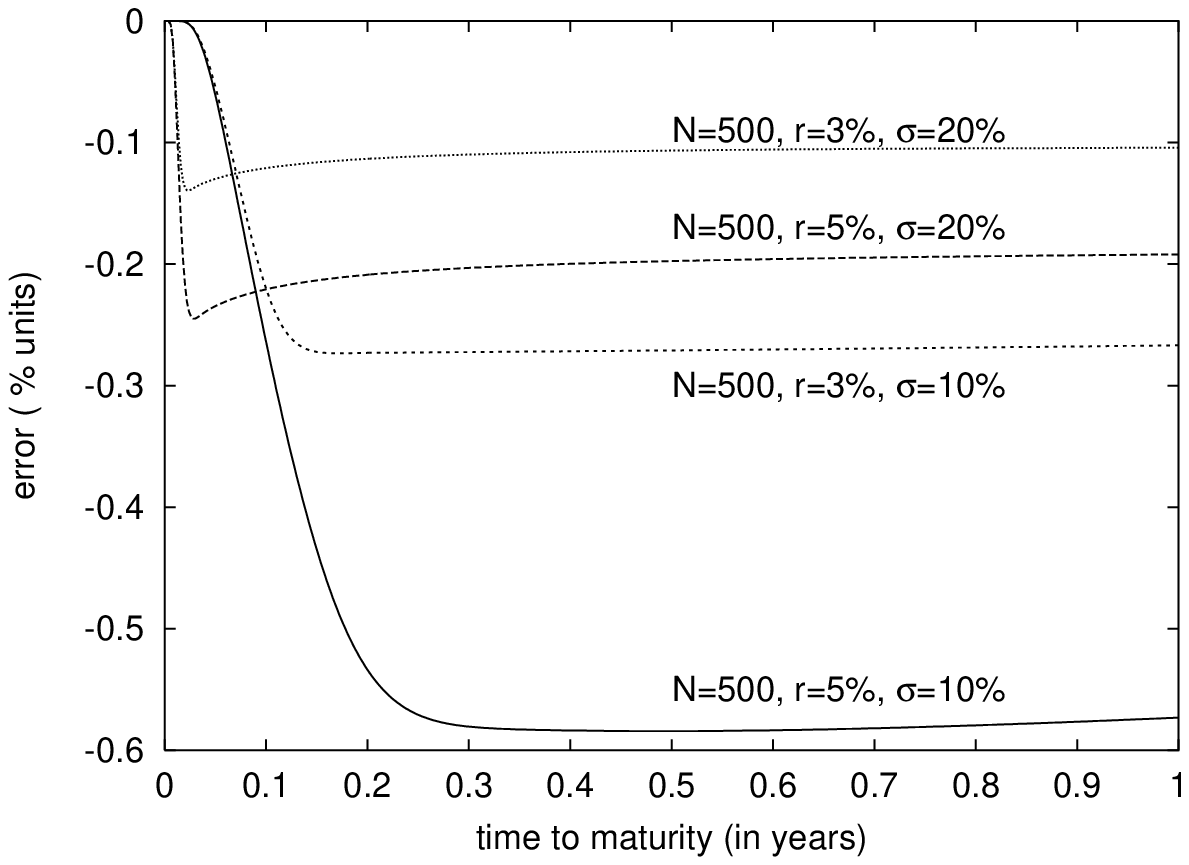,scale=0.54}
\end{center}
\caption{Computation of the put price derivative on the optimal exercise boundary and its resulting error in terms of $T-t$ when $K=100$. Left hand side plot shows numerical integration~(\ref{deri}) with the estimated values of $c(t)$ with a stepped function. This estimator of $c(t)$ has non constant derivative and differs from the theoretical value -1. Right hand side plot depicts error~(\ref{error}) quantifying the non-fulfillment of the smooth pasting condition when $S=95$.}
\label{fig2}
\end{figure}

Let us study this particular point in more detail. First we insisit that our interest is not the efficiency of the computation but the goodness of the price formula~(\ref{american}) based on the smooth pasting condition. We focus on the particular case given by Ref.~\cite{kim} where we use a step function to estimate the theoretical optimal exercise boundary $c(t)$. Equation~(\ref{optimal}) is easily solved by the Newton's method due to the monotonicity of the boundary curve. Figure~\ref{fig1} shows the rapid convergence of the estimated boundary curve and also depicts the results for several values of $r$ and $\sigma$. Once we get an estimation of the optimal exercise boundary, we calculate the American option price given by Eq.~(\ref{american}). 

And, what about the smooth pasting condition? From Eq.~(\ref{american}), we can calculate the derivative with respect to the stock and evaluated when $S>c(t)$:
\begin{eqnarray}
p_S^A(t,S)=p_S^E(t,S)-\frac{rK}{S}\int_t^T \frac{e^{-r(u-t)}}{\sigma\sqrt{u-t}}N'[d_2(S,d(u),u-t)]du,
\label{deri}
\end{eqnarray}
where $p_S^E(t,S)=-N[-d_1(S,K,T-t)]$. This derivative, popularly called delta, is a very important quantity since it prescribes the hedging strategy to be followed. In case we deal with the optimal exercise boundary $c(t)$, the smooth pasting condition~(\ref{smooth}) prescribes that this derivative is -1. However, in case we take our estimation of $c(t)$ represented by $S=d(t)$ and numerically calculate Eq.~(\ref{deri}), we find that our estimation does not obey the smooth pasting condition. Figure~\ref{fig2} shows that the numerical values of Eq.~(\ref{deri}) are different from -1. This questions the validity of the numerical computation and suggest a careful investigation (see Ref.~\cite{bermin}).

Having this inconsistency in mind, we provide a more general framework where not only optimal curves are allowed. Let us denote by $\mathcal{C}_{0,T}$ the set of non-decreasing continuous functions $d:\left[0,T\right] \rightarrow (0,K]$ with $d(T)=K$. Hence, the set $\mathcal{C}_{0,T}$ consists of all candidates for the optimal boundary. Consequently, following the spirit of Eq.~(\ref{americandef}), the price of the American put can be written as
\begin{equation}
p^{A}(t,S) =\sup_{d\in \mathcal{C}_{0,T}}\left\langle e^{-r(\tau_{t,d}-t)}[K-S(\tau)]^{+}|S(t)=S\right\rangle\equiv\sup_{d\in \mathcal{C}_{0,T}} p^{d}(t,S),
\label{newdef}
\end{equation}
where $\tau _{t,d}=\inf\{s\geq t:S(s)\leq d(s)\} \wedge T$. In Ref.~\cite{bermin}, we have analyzed the time $t$ price $p^{d}(t,S)$ for a general boundary $d\in \mathcal{C}_{0,T}$. The most important result found is that determinsitic function $p^d$ can be expressed as
\begin{eqnarray}
p^{d}(t,S)&=&p^{E}(t,S)+rK\int_{t}^{T}e^{-r(u-t)} N\left[-d_2(S,d(u),u-t)\right]du \nonumber \\
&-&\frac{\sigma}{2}\int_{t}^{T}\frac{e^{-r(u-t)}}{\sqrt{u-t}}d(u)N'\left[-d_2(S,d(u),u-t)\right]\left(1+p_{x}^{d}\left(
u,d(u\right)^+)\right)du,\nonumber\\
\label{extension}
\end{eqnarray}
where prime in the third term denotes the derivative $dN(x)/dx$. In order to derive this extension of the early exercising premium formula for an American put, we have required $p^d$ to be the solution of the PDE~(\ref{pde}) with conditions Eqs.~(\ref{below}) and~(\ref{smooth}). Details of this calculation are given in Ref.~\cite{bermin}. The expressions~(\ref{pde})--(\ref{smooth}) are essential new ingredients and, in some sense, they introduce a third term containing the partial derivative over the stock and evaluated on the curve $d(t)$. Note that, in case we deal with the optimal exercise boundary curve $c(t)$, the smooth pasting condition make this term to be zero and our extension of the early exercising premium formula would collapse to the American option put price (cf. Eq.~(\ref{american})). Hence, intuitively, the second integral in Eq.~(\ref{extension}) is able to quantify the error committed in computations aiming to get a price for the American put and reads
\begin{eqnarray}
\mbox{Error}\equiv-\frac{\sigma}{2}\int_{t}^{T}\frac{e^{-r(u-t)}}{\sqrt{u-t}}d(u)N'\left[-d_2(S,d(u),u-t)\right]\left(1+p_{x}^{d}\left(
u,d(u\right)^+)\right)du.
\nonumber\\
\label{error}
\end{eqnarray}
This quantity is directly related to the smooth pasting condition and lets us go beyond in judging the quality of an estimation of the optimal boundary based on the number of steps in the time discretization since, as we have seen in left hand side plot of Fig.~\ref{fig2}, even for a large number of steps the smooth pasting condition may not be accomplished. From Eqs.~(\ref{extension})-(\ref{error}), we thus take $(\mbox{Error}/p^d)\times 100$ and represent it in the right hand side plot of Fig.~\ref{fig2}. We have done it in terms of the maturity time, for several values of $r$ and $\sigma$, and when $S=95$. We observe that error is negative and non-negligible making always {\it cheaper} the corrected American put.

In this paper, we have approached to computational pricing problem using tools from mathematical finance. From these investigations, we may assert that the important but ignored properties of the American put option distort numerical results obtained via the binomial pricing method. We have focussed on the simplest numerical technique that takes the optimal exercise boundary curve as a step function and showed that the resulting price does not accomplish the smooth pasting condition. We have proposed an extension of the early exercise premium formula that captures the effect of the non-fulfilment of the smooth pasting condition. We have observed that errors are non-negligible and, most importantly, they can be quantified analytically. For this reason, we believe that the extended formula~(\ref{extension}) can provide fruitful results with a good mathematical ground as to ascertain whether a numerical method works well and it is thus able to evaluate the quality of existing computational methods in the literature. For more results we refer to the investigation in Ref.~\cite{bermin}.

\ack
Perell\'o wants to acknowledge the financial support by Direcci\'on General de
Investigaci\'on under contract No. BFM2003-04574 and by Generalitat de Catalunya under contract No. 2001 SGR-00061, Kohatsu-Higa the DGES grant and Bermin the support received from the Wallander Foundation.

\end{document}